\documentclass[%
reprint,
amsmath,amssymb,
aps,
prb,
]{revtex4-2}

\usepackage{graphicx}
\usepackage{epstopdf}
\usepackage{subfigure}
\usepackage{pifont}
\usepackage{bm}
\usepackage{xcolor}
\usepackage{tikz}

\usepackage[colorlinks=true,linkcolor=blue,anchorcolor=blue,citecolor=blue,urlcolor=blue]{hyperref}
\usepackage[utf8]{inputenc} 
\DeclareUnicodeCharacter{200B}{}
\DeclareUnicodeCharacter{202F}{ }
\begin{document}

\preprint{APS/123-QED}

\title{Sublattice polarization and filamentary superconductivity in strained graphene
}

\author{Tao Zhou\textsuperscript{1,2}}
\email{tzhou@scnu.edu.cn}
\affiliation{ \textsuperscript{1}Guangdong Basic Research Center of Excellence for Structure and Fundamental Interactions of Matter, \\
Guangdong Provincial Key Laboratory of Quantum Engineering and Quantum Materials, \\
School of Physics, South China Normal University, Guangzhou 510006, China \\
\textsuperscript{2}Guangdong-Hong Kong Joint Laboratory of Quantum Matter, \\
Frontier Research Institute for Physics, South China Normal University, Guangzhou 510006, China}

\begin{abstract}

Periodic strain engineering in monolayer graphene provides a versatile platform to generate colossal pseudo-magnetic fields and flat pseudo-Landau levels directly in a single atomic sheet. While the electronic topology and transport properties of the normal state in these systems have been extensively studied, the superconducting ground state in such strain-modulated landscapes remains an unexplored frontier. In this work, we investigate the superconducting phase of periodically corrugated graphene using a self-consistent Bogoliubov-de Gennes framework. We find that, contrary to the conventional expectation that a high density of states universally enhances pairing, the macroscopic superconducting coherence is significantly hindered in the flat-band regions. This limitation originates from strain-induced sublattice polarization, where the zero-energy electronic states are spatially segregated between the $A$ and $B$ sublattices. Such spatial disjointedness hampers the inter-sublattice coherence required for a robust pairing instability, effectively decoupling the high-density flat-band states from the superconducting condensate. Consequently, as the pairing interaction increases, the system undergoes a striking spatial crossover: superconductivity sharply relocates from the flat-band regions to emerge as robust, quasi-one-dimensional filaments at the geometric nodes where local sublattice symmetry is restored. Our findings reveal that the spatial distribution of wavefunctions, governed by sublattice degrees of freedom, is a decisive factor in determining the superconducting properties of strain-engineered Dirac materials.
\end{abstract}

\maketitle

\section{introduction}

Since the isolation of graphene, realizing intrinsic superconductivity has been a central goal in condensed matter physics. The primary obstacle is the vanishing density of states (DOS) at the Dirac point. Early strategies primarily focused on extrinsic modifications, such as heavy chemical doping ~\cite{Ichinokura2016,Chapman2016,doi:10.1073/pnas.1510435112}, substrate interaction~\cite{PhysRevLett.111.246805}, or proximity-induced pairing~\cite{DiBernardo2017,Lee_2018,adma.202008113}, which, while effective, inherently alter or mask the pristine nature of the carbon lattice.

A paradigm shift occurred recently with the rise of crystalline engineering in multilayer systems. Remarkable superconductivity has been achieved in Bernal-stacked bilayer graphene~\cite{doi:10.1126/science.abm8386,Zhang2023}, rhombohedral trilayer graphene~\cite{Zhou2021}, and most notably, magic-angle twisted bilayer and trilayer graphene~\cite{Cao2018,Cao2021,Park2021,Park2022,science.abn8585,science.abg0399,science.adv8376}. These breakthroughs share a common principle: manipulating interlayer stacking or twist angles to engineer the electronic structure. Particularly in magic-angle twisted systems, the twist angle acts as a critical tuning parameter to engineer flat bands and van Hove singularities. In this flat-band regime, the resulting divergent DOS is widely regarded as the primary driver for strong correlations and superconductivity.

Structurally deformed graphene offers a parallel geometric route, distinct from the multi-layer twisting approach~\cite{Meyer2007,Fasolino2007}. Strain engineering can generate colossal pseudo-magnetic fields (PMF) and flat pseudo-Landau-levels (PLLs) directly in a single atomic sheet~\cite{PhysRevLett.101.226804,PhysRevB.77.075422,Wehling_2008,PhysRevB.81.115421,VOZMEDIANO2010109,Guinea2010,PhysRevB.87.205405,Xu2024,PhysRevB.92.245302,PhysRevLett.129.056803}. Although the electronic topology and valley-polarized transport in the normal state of these systems have been extensively studied, ~\cite{PhysRevLett.101.226804,PhysRevB.77.075422,Wehling_2008,PhysRevB.81.115421,VOZMEDIANO2010109,Guinea2010,PhysRevB.87.205405,Xu2024,PhysRevB.92.245302,PhysRevLett.129.056803}, the superconducting ground state in such strain-engineered landscapes remains an entirely unexplored frontier. This investigation is particularly crucial: since strain shares the capability of twist to flatten electronic bands, does the high DOS in strain-engineered graphene enhance superconductivity as it does in twisted systems?

In this paper, we address this question by investigating the superconducting phase of periodically corrugated graphene using a self-consistent Bogoliubov-de Gennes framework. Contrary to the common intuition that a high DOS universally promotes pairing, we identify a striking interaction-driven spatial crossover of the superconducting order parameter governed by strain-induced sublattice polarization.
In the flat-band regions, the wavefunctions of the $n=0$ PLLs are localized almost exclusively on a single sublattice (A or B), with a marked spatial segregation between the two. This geometric disjointedness effectively hampers the inter-sublattice overlap required for a coherent condensate, leading to a suppression of the pairing amplitude despite the divergent DOS.
Consequently, we find that robust superconductivity selectively emerges not in the high-DOS regions, but as quasi-one-dimensional filaments at the geometric nodes where local sublattice symmetry is restored. Furthermore, we demonstrate that non-magnetic impurities in this highly inhomogeneous state induce resonant peaks at the gap boundaries due to strong quasi-one-dimensional confinement. These unique gap-edge resonances provide a distinctive local probe for identifying the filamentary superconducting nature in strain-engineered Dirac materials.

The remainder of this paper is organized as follows: Section~II presents the model and the formalism; Section~III discusses the numerical results; and Section~IV provides a summary.

\section{MODEL AND FORMALISM}
We consider a superconducting system with strain-modulated hopping.
The normal state Hamiltonian is given by the tight-binding model:
\begin{equation}
    H_N = -\sum_{\langle i,j \rangle, \sigma} t_{ij} c_{i\sigma}^\dagger c_{j\sigma} 
    - \mu \sum_{i,\sigma} c_{i\sigma}^\dagger c_{i\sigma},
    \label{eq:H0}
\end{equation}
where $t_{ij}$ denotes the hopping integral between sites $i$ and $j$, $\mu$ represents the chemical potential, $c_{i\sigma}^\dagger$ and $c_{i\sigma}$ are the creation and annihilation operators for an electron with spin $\sigma$ at site $i$, respectively, and $\langle i,j \rangle$ indicates summation over nearest-neighbor pairs.


The hopping integral is modulated by unidirectional strain, corresponding to a periodically corrugated structure along the 
$x$-direction while remaining uniform along 
$y$. The corrugation is modeled as a sinusoidal ripple:
\begin{equation}
    z(x) = h \sin\left( \frac{2\pi x}{L} \right),
    \label{eq:ripple}
\end{equation}
where $H$ is the corrugation amplitude and $L$ is its period along $x$.

The strain-renormalized hopping integral follows the form
\begin{equation}
    t_{ij} = t_0 \exp\left[ -\beta \left( \frac{d_{ij}}{a_0} - 1 \right) \right],
    \label{eq:tij_strain}
\end{equation}
where $t_0$ is the unstrained nearest-neighbor hopping amplitude. $\beta$ is the decay constant with $\beta=3.37$~\cite{PhysRevB.80.045401,PhysRevB.81.081407}. $d_{ij}$ denotes the strained distance between sites $i$ and $j$, and $a_0$ is the equilibrium carbon-carbon bond length in graphene.

We now consider the superconducting pairing term. Microscopically, electron-phonon interactions, as described by BCS theory, can mediate superconducting pairing. Such pairing occurs within a thin momentum-space shell around the Fermi surface. At the mean-field level, the required attractive interaction can be rather weak. In real space, the superconducting coherence length $\xi$ is typically large, and the pairing function is a superposition of many $n$th neighbor contributions ($n=0,1,2,\dots$) up to the order of $\xi$.

In a translationally invariant system, the conventional $s$-wave pairing is isotropic in momentum space. One common real-space approximation is to use a phenomenological on-site attractive interaction $V$, which has been shown to effectively mimic the essential physics of $s$-wave pairing \cite{PhysRevB.93.094517}. However, a crucial difference is that on-site pairing couples electrons across the entire Brillouin zone, making it a strong-coupling interaction. Consequently, a finite threshold pairing strength $V_c$ is required to induce superconductivity, whereas the BCS-type electron-phonon pairing is weak-coupling and has no such threshold. Despite this difference, the geometric effects arising from the strain-modulated lattice, which are the main focus of this work, do not depend on the detailed pairing mechanism as long as the pairing is spin-singlet and requires inter-sublattice coherence.

In the superconducting state, we therefore adopt a phenomenological on-site attractive interaction $V$ as a minimal model for conventional $s$-wave pairing. This choice allows us to isolate the geometric effects of the strain-induced gauge fields from the complexities of unconventional pairing symmetries. The Hamiltonian is diagonalized by solving the BdG equation,
\begin{equation}
\sum_j 
\begin{pmatrix} 
-t_{ij} - \mu\delta_{ij} & \Delta_{ij} \delta_{ij} \\ 
\Delta_{ij}^* \delta_{ij} & t_{ij}^* + \mu\delta_{ij} 
\end{pmatrix} 
\begin{pmatrix} 
u_j^n \\ v_j^n 
\end{pmatrix} 
= E_n 
\begin{pmatrix} 
u_i^n \\ v_i^n 
\end{pmatrix}.
\end{equation}

The superconducting order parameters at site $i$ are determined self-consistently at finite temperature $T$ via
\begin{equation}
 \Delta_{i} = \frac{V}{2} \sum_n u_{i,n} v_{i,n}^* \tanh\left( \frac{E_n}{2T} \right),
    \label{eq:Gap_finiteT}
\end{equation}
where $V$ denotes the attractive pairing interaction strength.

The local density of states (LDOS) at site $i$ and energy $E$ is computed by summing over all energy bands:
\begin{equation}
    \rho_i(E) = \sum_n\left[ |u_{i,n}|^2 \delta(E - E_{n}) + |v_{i,n}|^2 \delta(E + E_{n}) \right].
    \label{eq:LDOS_full}
\end{equation}
For numerical calculation, the delta function is approximated by a Lorentzian broadening:
\begin{equation}
    \delta(x) \approx \frac{1}{\pi} \frac{\Gamma}{x^2 + \Gamma^2},
    \label{eq:Lorentzian}
\end{equation}
where $\Gamma$ is a small broadening parameter representing finite lifetime effects.

It is worth emphasizing that our primary goal is to resolve the spatial texture of superconductivity induced by periodic strain. Advanced techniques like dynamical mean-field theory or Eliashberg theory, while powerful for dynamical correlations, are currently prohibitive for systems with large size. 
For systems with large supercells and significant real-space inhomogeneity, the self-consistent real-space BdG framework is the established and most practical method.
With this approach, 
we have performed self-consistent calculations with different system sizes along the $y$ direction, using a smaller $N_x$ and larger $N_y$ (up to $N_y = 30$). The results show that the order parameter is strictly uniform along $y$ for all tested widths. Hence, the system is uniform along $y$, and $k_y$ is a good quantum number.
Then we consider a unidirectional corrugated modulation with a period of 
$L=500$ unit cells
 and a corrugation ratio 
$r=h/L=0.16$, a value within experimental reach~\cite{Zang2013,Blees2015}.
We use the unstrained nearest-neighbor hopping $t_0$
  as the unit of energy. Other parameters are set as follows: the temperature 
$T=10^{-5}$, the chemical potential $\mu=0$, and the spectral broadening 
$\Gamma=0.004$.

\section{results and discussion}

Before elucidating the superconducting state, we first characterize the electronic landscape of the normal state. As illustrated in Fig.~1(a), the unidirectional corrugation breaks the translational symmetry, generating a periodic PMF. The resulting band structure [Fig.~1(b)] exhibits flat bands at the Fermi energy ($E=0$), corresponding to the $n=0$ PLLs. 
These flat bands originate from the low-energy Dirac points at the two inequivalent valleys, 
$K$ and $K'$. A defining feature of these strain-induced states, distinct from those generated by real magnetic fields, is the locking of the valley degree of freedom to the sublattice: due to the opposite signs of the PMF at the two valleys, the zeroth PLLs from the 
$K$ valley localize exclusively on one sublattice, while those from the 
$K'$ valley localize on the other.

\begin{figure}
\centering
  \raisebox{1.7cm}{
\begin{tikzpicture}[baseline=(current bounding box.center)]
 \node[anchor=south west, inner sep=0] (image) at (0,0) {
 \includegraphics[width=3.6cm]{{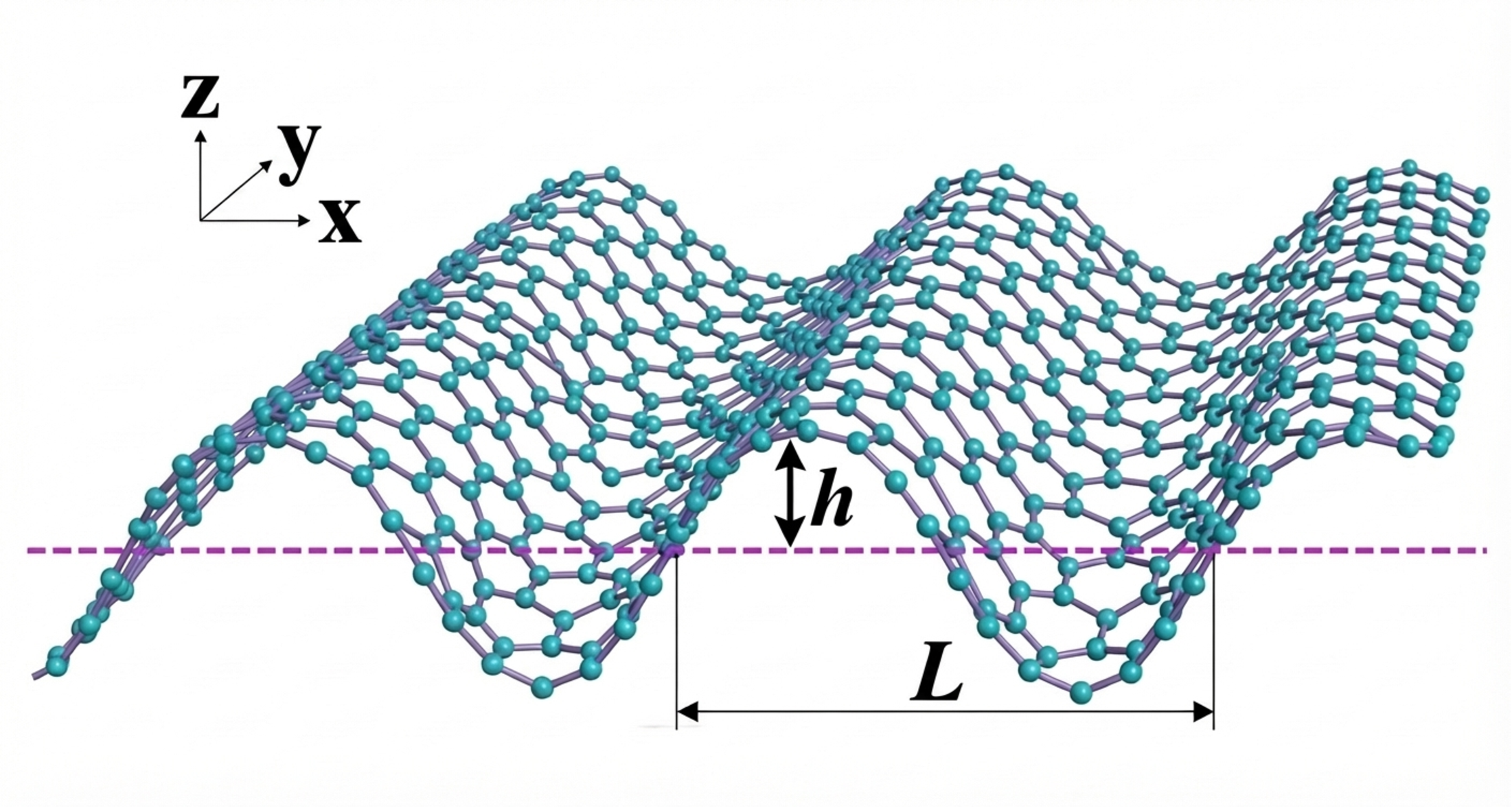}}  };
 \node[anchor=north west] at (0.26, 2.2) {\tiny \textsf{(a)}}; 
\end{tikzpicture}}           
       \includegraphics[width = 3.6cm]{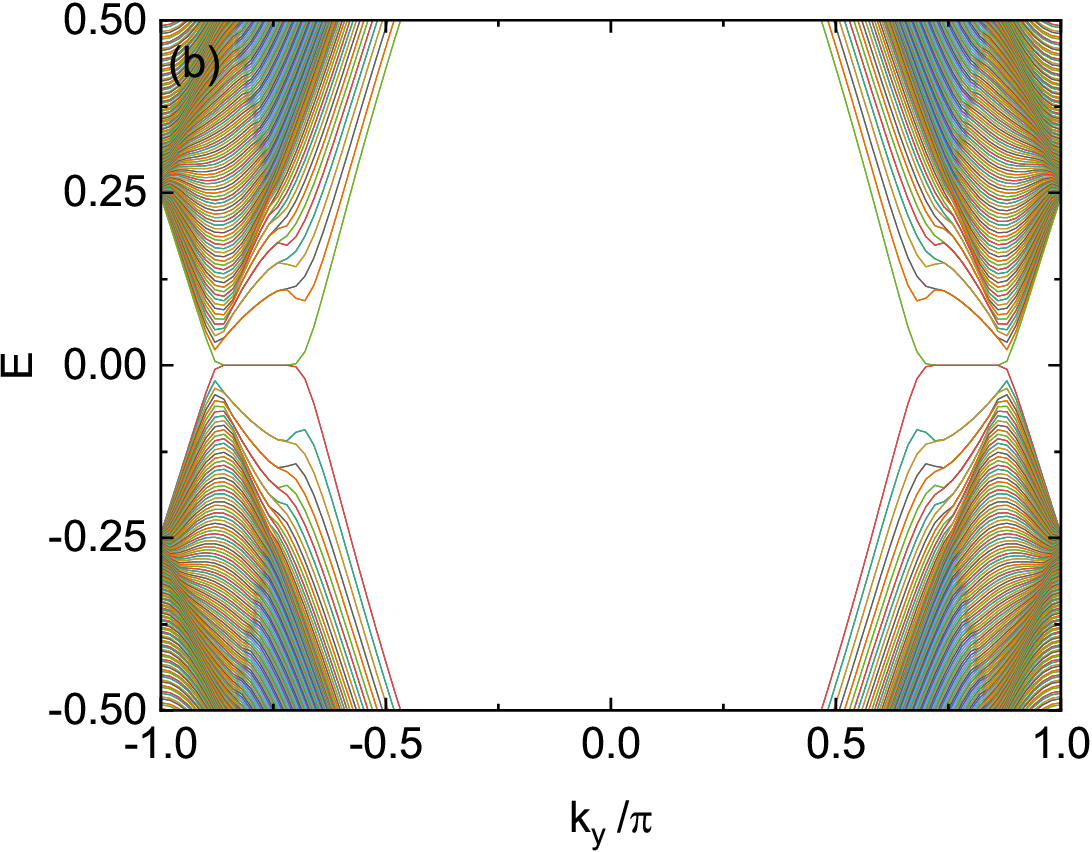}
        \includegraphics[width = 3.6cm]{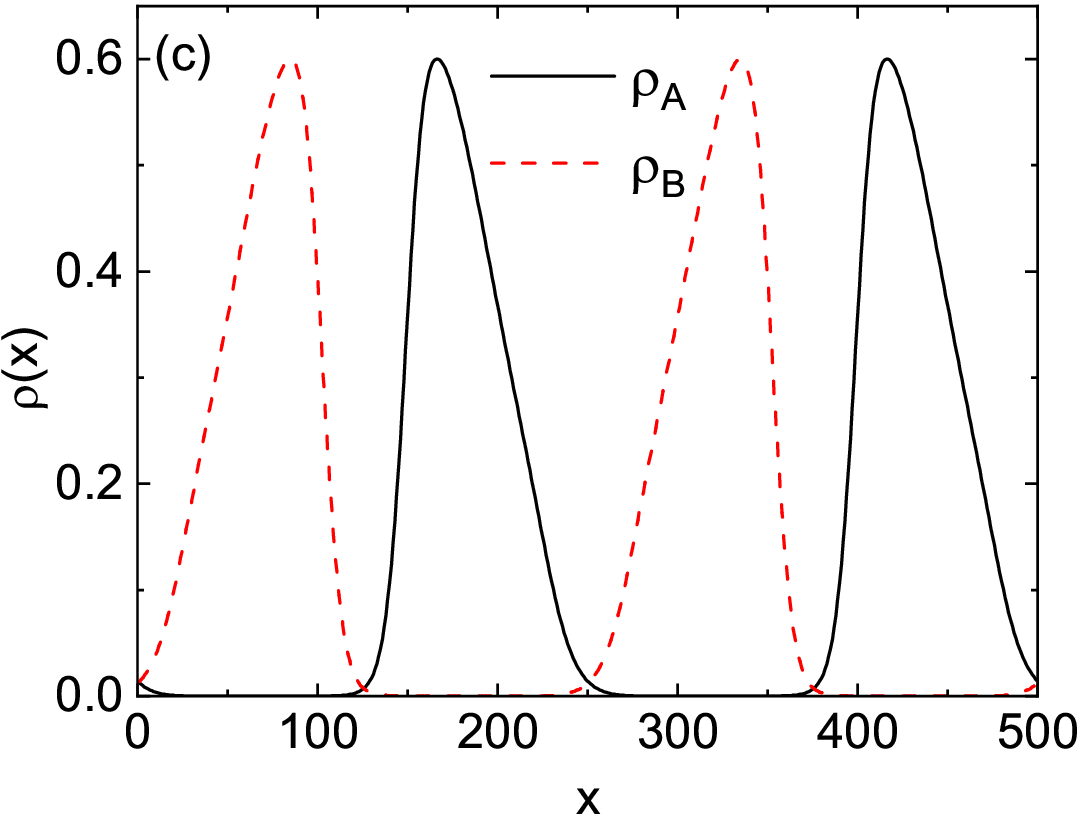}
        \includegraphics[width = 3.6cm]{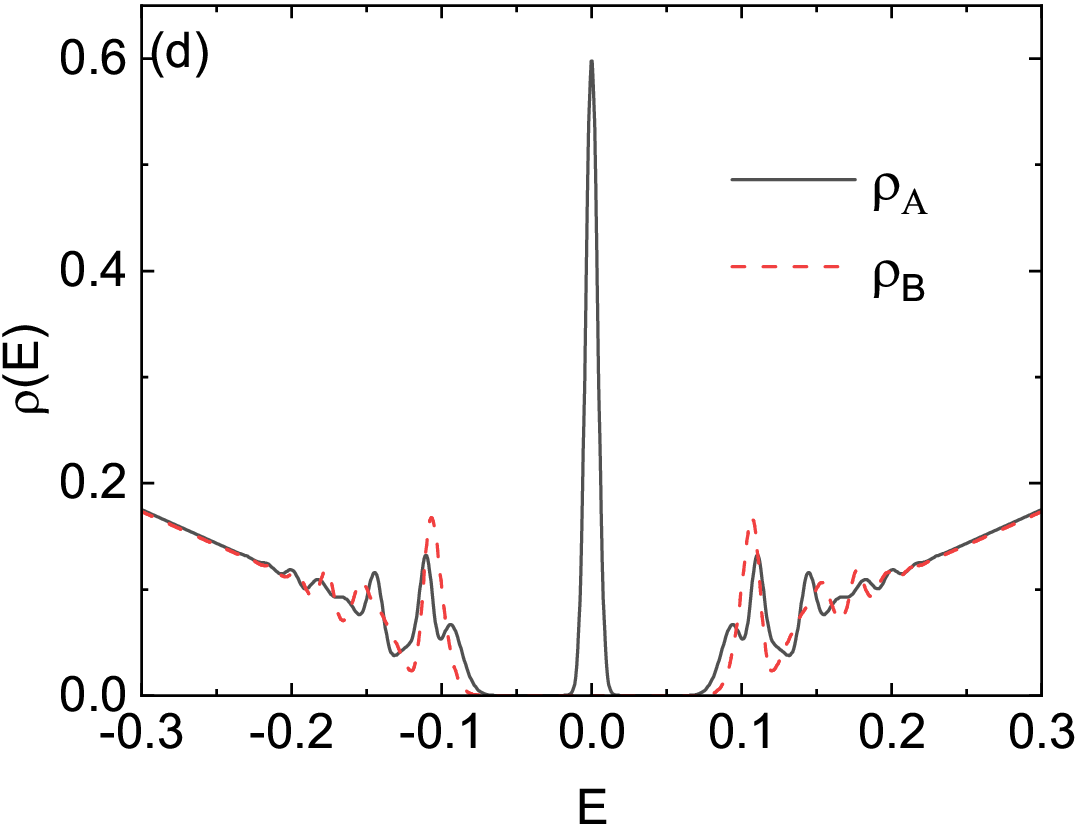}
	\caption{\label{fig1}Electronic structure of the corrugated graphene in the normal state. (a) Schematic illustration of the sinusoidally strain-engineered graphene lattice. (b) The calculated energy band structure along 
$k_y$​, exhibiting flat bands at zero energy induced by the PMF. (c) Spatial profile of the zero-energy LDOS along the corrugated direction 
$x$. The solid (dashed) lines represent the A (B) sublattice, revealing a distinct spatial separation of the zeroth PLLs. (d) Energy-dependent LDOS at 
$x=0.33L$, showing a sharp peak at the Fermi level for the A-sublattice (solid line) and a gap for the B-sublattice (dashed line). }
\end{figure}

This valley-dependent segregation is directly visualized in the zero-energy LDOS in Fig. 1(c): the wavefunctions are spatially separated, with A-sublattice states localized near $x=0.33L$ and 
$0.83L$, and B-sublattice states peaking at $x=0.17L$ and 
$0.67L$. This leads to a scenario where, locally, the high-DOS flat bands are hosted almost exclusively by a single sublattice [Fig.~1(d)]. This sublattice segregation is not merely a spectral feature but a fundamental obstruction to pairing, as we discuss below.

We now turn to the self-consistent solution of the superconducting order parameter $\Delta(x)$. Since the system remains translationally invariant along the
$y$ direction, the order parameter is uniform along $y$. Fig. 2 plots the spatial profile of the order parameter amplitude along the corrugation direction 
$x$ for different pairing interaction strengths $V$. Our calculations reveal a striking interaction-driven spatial crossover in the superconducting state, highly dependent on the magnitude of the pairing potential.

For a relatively weak pairing potential ($0.3 < V < 1.1$), i.e., just above the threshold $V_c \approx 0.3$, superconductivity is primarily nucleated within the PMF-induced flat-band regions. In this regime, the overall pairing amplitude is relatively small ($\sim 10^{-3}$). Crucially, the spatial profile exhibits a pronounced sublattice asymmetry, with $\Delta_A(x)$ and $\Delta_B(x)$ showing distinct spatial distributions. This asymmetry is a direct macroscopic manifestation of the sublattice polarization of the zero-energy localized states in the normal state [as shown in Fig.~1(c)], where the constrained wavefunction overlap limits the phase coherence between the $A$ and $B$ sublattices.

However, as the pairing potential increases, the system undergoes a dramatic spatial reconstruction. As illustrated in Fig.~2(b) for $V = 1.2$, the pairing amplitude in the flat-band regions is superseded by a sharp, massive enhancement of superconductivity at the geometric nodes of the corrugation (e.g., around $x = 0.5L$). At these nodal regions, the local sublattice symmetry is fully restored ($\Delta_A = \Delta_B$), and the pairing amplitude increases by more than an order of magnitude. This feature indicates the emergence of robust, quasi-one-dimensional superconducting filaments. In contrast, the pairing amplitude $\Delta$ in the flat-band regions remains small, contradicting the conventional expectation that a high DOS enhances superconductivity.

	\begin{figure}
	\centering        
	\includegraphics[width = 8cm]{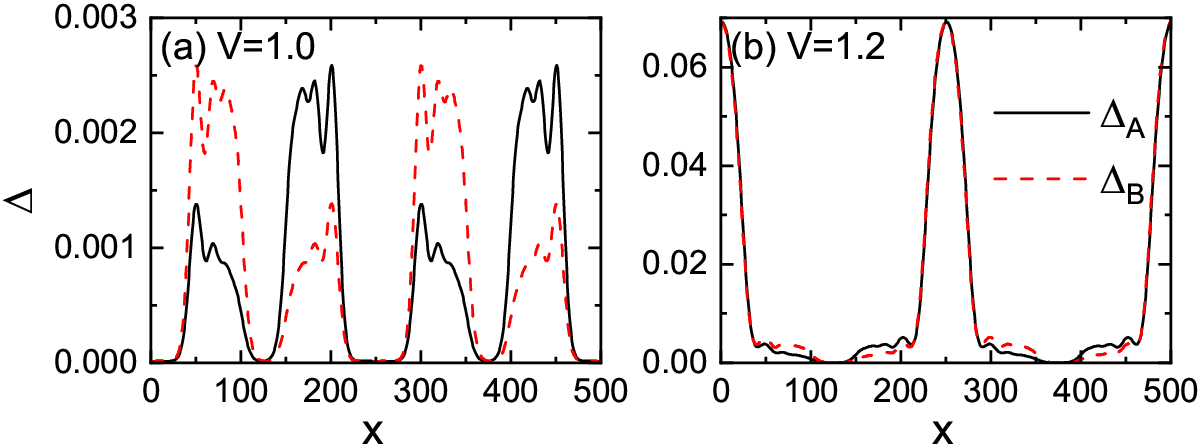}
	\caption{\label{fig2} Self-consistent profiles of the superconducting order parameter amplitude along the corrugation direction 
		$x$ for (a)  $V=1.0$ and (b) $V=1.2$.  }
\end{figure}

The suppression of the pairing amplitudes at the flat band regions is a direct consequence of the extreme sublattice polarization induced by the PMF. In these regions, the electronic density of the zeroth PLLs is localized almost exclusively on a single sublattice while vanishing on the other [Figs. 1(c) and 1(d)]. Such spatial segregation effectively severs the nearest-neighbor wavefunction overlap between the $A$ and $B$ sublattices.
Consequently, the superconducting state inherits this polarization, manifesting as a marked asymmetry between the sublattice order parameters ($\Delta_A\neq\Delta_B$), as evident in Fig. 2. Since the establishment of a coherent 
$s$-wave condensate relies on the effective kinetic mixing and overlap between sites to lower the free energy, this spatial disjointedness decouples the localized flat-band states from the pairing instability. This behavior highlights a fundamental physical distinction between the strain-induced flat bands and the moiré flat bands in magic-angle twisted bilayer graphene (MATBG).

 In MATBG, it has been established that while flat bands traditionally imply strongly localized states or divergent effective masses that would suppress superconductivity~\cite{Liu_2014}, the quantum geometry of the Bloch wavefunctions-characterized by the quantum metric-provides an additional, non-zero contribution to the superfluid weight~\cite{Torma2022}. This geometric contribution allows Cooper pairs to remain delocalized and establish phase coherence even when the single-particle bandwidth vanishes.

In contrast, the periodically strained graphene studied here appears to lack such geometric protection. The extreme sublattice polarization in the 
$n=0$ PLL regions leads to a real-space segregation of wavefunctions, which effectively quenches inter-site overlaps and prevents the formation of a non-trivial quantum metric that would otherwise facilitate a finite superfluid weight. Moreover, the strain-induced PMF is valley-contrasting, and the absence of robust inter-valley coupling in these localized regions further hinders the topological stabilization of the condensate~\cite{PhysRevLett.61.2015}. Finally, unlike MATBG where flat bands span the entire moiré Brillouin zone, the strain-induced flat bands shown in Fig. 1(b) are flat only in a limited momentum-space region, further limiting the potential for a global geometric contribution to superconductivity.

The geometric obstruction, which effectively decouples the localized wavefunctions in real space, would likely be even more detrimental to off-site pairing symmetries (e.g., 
$d$-wave) that rely explicitly on inter-site bonds for Cooper pair formation. Therefore, our model provides a conservative estimate of the disruptive impact that strain-induced sublattice polarization has on superconductivity. It suggests that the high density of states in strain-induced flat bands does not necessarily facilitate pairing, but instead leads to a spatial dissociation of the condensate due to the underlying sublattice degrees of freedom.

Conversely, at the nodes of corrugation ($z=0$), the physics is governed by bond stretching rather than the PMF. Here, the vanishing PMF ensures the restoration of A-B sublattice symmetry, facilitating the necessary wavefunction overlap. 
The maximal tensile strain exponentially suppresses the local hopping $t_{ij}$ [Eq.~(2)], thereby enhancing the effective correlation ratio $V/t_{\text{eff}}$. This drives the system locally into a strong-coupling regime. Unlike the fragile flat-band states, the electrons at the nodes retain sufficient A-B sublattice overlap to support robust phase coherence, resulting in the formation of quasi-one-dimensional superconducting filaments.

Additionally, we observe that $\Delta_{x}$ vanishes at the crests and troughs ($x=0.25L$ and 
$0.75L$). Although the A-B sublattice symmetry is fully recovered in these zero-PMF regions, the local slope of the sinusoidal corrugation profile $z(x)$ [Eq. (2)] vanishes, meaning the lattice distortion is minimal. Consequently, the effective hopping 
$t_{eff}$ relaxes back to the unstrained value 
$t_0$​. Lacking both the flat bands and the strain-induced bandwidth narrowing, these regions effectively mimic pristine graphene, which remains in the weak-coupling limit and is non-superconducting at the simulated interaction strength.

As shown in Fig.~1(d), the normal-state LDOS at the Fermi level exhibits a sharp zero-energy peak, arising from the $n=0$ PLLs. When $\mu \neq 0$, this peak is immediately depleted, and the normal-state density of states at the Fermi energy becomes negligible. Consequently, the sublattice polarization effect disappears, and the pairing in the flat-band regions is no longer present. In contrast, the filamentary superconductivity at the geometric nodes is enhanced due to an increased normal-state density of states at the Fermi energy upon doping.

	\begin{figure}
		\centering        
		\includegraphics[width = 8cm]{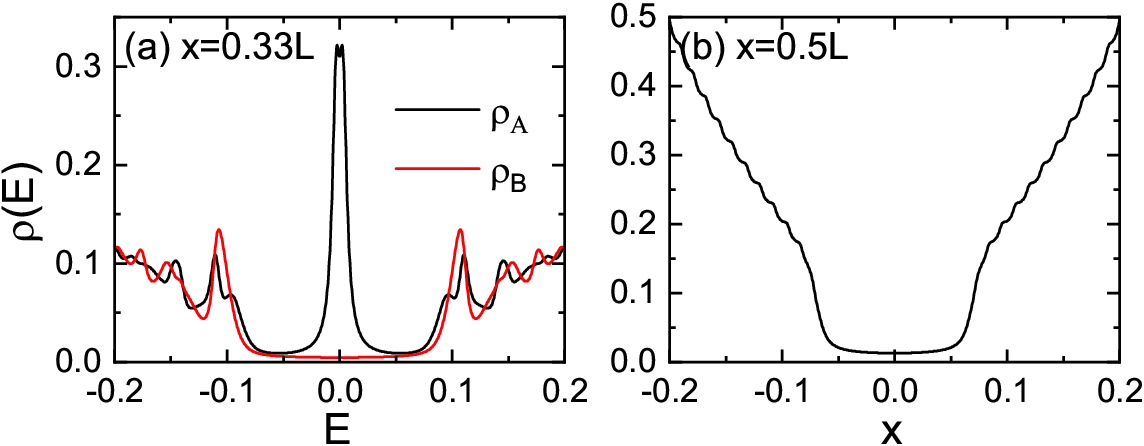}
		\caption{\label{fig3} LDOS in the superconducting state. (a) LDOS in the flat-band region ($x=0.33L$). The strong sublattice polarization persists (dashed vs. solid lines), but the original normal-state zero-energy peak is split, indicating the opening of a partial superconducting gap. (b) LDOS at the node of corrugation ($x=0.5L$), where the sublattice symmetry is restored. }
	\end{figure}

To further elucidate the microscopic nature of this highly inhomogeneous superconducting state, we calculate the LDOS
 at different spatial locations. Fig. 3 contrasts the LDOS in the flat-band region with that at the geometric node.

In the flat-band region ($x=0.33L$, Fig. 3a), the LDOS retains the strong sublattice polarization characteristic of the normal state, with the A and B sublattices remaining spectrally separated. However, a crucial change occurs: the sharp zero-energy peak observed in the normal state (Fig. 1d) is now split into two peaks centered away from the Fermi level. This splitting signifies the opening of a superconducting gap, 
$\Delta_{flat}$, on the flat bands. The persistence of sublattice separation implies that any Cooper pairs formed in this region must involve electrons residing on distinct sublattices that are spatially segregated. This severe spatial mismatch restricts the development of robust phase coherence, explaining why the local pairing amplitude remains weak despite the highly divergent normal-state DOS.

In stark contrast, at the nodes of the corrugation ($x=0.5L$, Fig. 3b), where the order parameter is maximized, the A and B sublattice spectra become identical, indicating a full restoration of local sublattice symmetry. A large, hard gap is clearly visible, reflecting the robust pairing amplitude within these quasi-one-dimensional filaments. Notably, the spectrum exhibits a striking absence of sharp coherence peaks at the gap edges, which are typical hallmarks of conventional BCS superconductivity. This suppression of coherence peaks is a direct consequence of the extreme spatial inhomogeneity of the order parameter. The effective reduction of dimensionality—confining the superconducting condensate to quasi-1D filaments—strongly redistributes the spectral weight, leading to a smooth gap profile rather than the sharp spectral pile-ups expected in uniform two-dimensional systems.

To probe the unique properties of the filamentary superconducting state, we investigate its local electronic response to a single non-magnetic impurity. We introduce a localized impurity potential $H_{\text{imp}} = V_i \sum_{\sigma} c_{0\sigma}^\dagger c_{0\sigma}$ at a specific site within the nodal superconducting filament ($x=0.5L$) and monitor the evolution of the LDOS at its nearest-neighbor site.

\begin{figure}
		\centering        
		\includegraphics[width = 7cm]{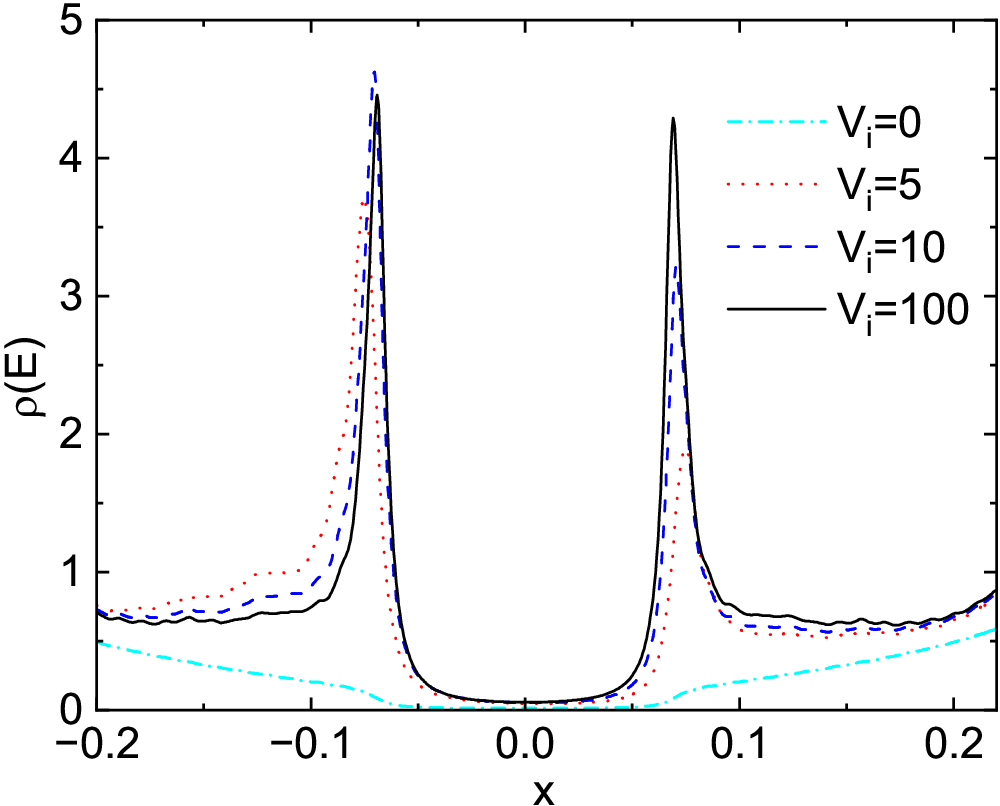}
		\caption{\label{fig4}
Impurity-induced gap-edge resonance in the filamentary superconducting state. The LDOS at a nearest-neighbor site of a non-magnetic impurity located in the superconducting region is plotted for various impurity strengths $V_i$.  }
	\end{figure}

Fig. 4 illustrates the impurity-induced spectral variations for different scattering strengths 
$V_i$​. As the impurity strength increases, no localized in-gap states or zero-bias conductance peaks are induced. Instead, intense, highly localized resonant peaks emerge exactly at the gap boundaries ($E\approx \pm\Delta$). As $V_i$​ approaches the unitary limit ($V_i​=10$), these gap-edge peaks become exceptionally sharp and exhibit particle-hole symmetry.

This unconventional impurity response is driven by the quasi-one-dimensional confinement effect inherent to the filamentary superconductivity. In this quasi-one-dimensional geometry, a strong non-magnetic impurity acts as a hard-wall scattering barrier. Quasiparticles undergo multiple normal and Andreev reflections between the impurity site and the surrounding spatially varying pairing field. Due to the conventional 
$s$-wave nature of the local pairing, low energy in-gap bound states are forbidden~\cite{ref36,PhysRevB.111.174525}; consequently, these geometry-confined resonance modes are entirely pushed to the gap edges. The emergence of these giant, symmetric gap-edge resonance peaks under strong impurity scattering serves as a distinctive local signature of the quasi-one-dimensional filamentary confinement in corrugated graphene.

The periodic strain textures modeled here ($L = 500$ unit cells, corrugation ratio $r = 0.16$) are well within the reach of current strain-engineering capabilities. Large-amplitude corrugations and crumpled structures exceeding this aspect ratio have been realized by relaxing graphene on pre-stretched elastomeric substrates~\cite{Zang2013} or through kirigami-inspired mechanical manipulation~\cite{Blees2015}. These experiments demonstrate that monolayer graphene possesses exceptional mechanical resilience, sustaining high-curvature deformations without fracture. Consequently, these strain-engineered platforms provide an ideal arena to experimentally verify the predicted geometric dissociation of superconductivity and the associated filamentary superconductivity.

Our findings offer a timely theoretical perspective for interpreting the role of strain in correlated oxides, particularly in light of the very recent realization of ambient-pressure superconductivity in La$_3$Ni$_2$O$_7$ ​ thin film~\cite{Zhou2025, Ko2025}. In these experiments, epitaxial strain serves as a critical tuning parameter, substituting for the need for high physical pressure~\cite{ref1}. Notably, transport measurements suggest a possible filamentary or domain-wall character for the resulting superconducting state~\cite{Ko2025}. This phenomenology closely aligns with our predictions for corrugated graphene, where periodic strain drives a spatial dissociation of electronic states, confining superconductivity to specific geometric loci (the nodes) while strongly suppressing it elsewhere.

We emphasize that this comparison is intended as a phenomenological analogy rather than a direct theoretical mapping. Nevertheless, we propose that this strain-induced filamentation shares a fundamental phenomenological similarity across these physically distinct systems: the strain-driven disentanglement of internal degrees of freedom. In nickelates, the minimal model involves two active orbitals ($d_{z^2}$ and $d_{x^2-y^2}$), and theoretical studies suggest superconductivity is governed by an orbital-selective mechanism~\cite{PhysRevLett.131.126001,arXiv2311.05491,rm9g-8lm1}. Our corrugated graphene system presents a compelling geometric analog: the two sublattices ($A$ and $B$) play a role mathematically and physically analogous to these distinct orbitals.

The suppression of pairing we identify in the flat-band regions is essentially driven by a sublattice-selective spatial segregation. Just as strain can differentially tune orbital overlaps and hybridizations in nickelates, the PMF in corrugated graphene spatially segregates the wavefunctions of the $A$ and $B$ sublattices. Our work thus points to a generalized principle for strain-engineered superconductors: non-uniform strain does not merely renormalize electronic bandwidths, but acts as a spatial “phase filter.” It naturally segregates regions capable of supporting coherent pairing (the nodes) from regions where spatial disjointedness suppresses the condensate (the flat bands), universally leading to the emergence of filamentary or spatially textured superconductivity.

\section{summary}
In summary, we have investigated the superconducting phase of periodically strained monolayer graphene. Our results demonstrate an interaction-driven spatial crossover of superconductivity, fundamentally governed by strain-induced sublattice polarization. Although the periodic strain generates high-density-of-states flat bands (zeroth pseudo-Landau levels), these states are localized on individual sublattices with minimal spatial overlap between $A$ and $B$ sites. This geometric segregation inhibits the formation of a robust, coherent superconducting condensate in the flat-band regions at weak coupling. Instead, as the pairing potential increases, superconductivity sharply relocates and emerges as robust, quasi-one-dimensional filaments at the geometric nodes where local sublattice symmetry is restored. Furthermore, we demonstrate that a strong non-magnetic impurity in this highly inhomogeneous state generates resonant peaks precisely at the gap boundaries. We propose that these gap-edge resonances, arising from strong quasi-one-dimensional confinement and Andreev reflection, can serve as a definitive local experimental signature of the filamentary superconducting state. Our work highlights that in strain-engineered Dirac materials, the sublattice degree of freedom plays a decisive role in governing the superconducting ground state, offering a perspective distinct from other flat-band systems.

\bibliography{rip-graphene}

\end{document}